\documentclass[sigconf]{acmart}
\usepackage[spanish]{babel}
\AtBeginDocument{%
  }
\setcopyright{acmcopyright}
\copyrightyear{2023}
\acmYear{2023}
\acmDOI{XXXXXXX.XXXXXXX}

\acmConference[Interacción '23]{XXIII International Conference on Human Computer Interaction}{Septiembre 4--6, 2023}{Lleida, Spain}
\acmPrice{15.00}
\acmISBN{978-1-4503-XXXX-X/18/06}

\begin{document}

\title{Estudio de la Experiencia de Usuario mediante un Sistema de Dashboards de Análisis de Aprendizaje Multimodal}


\author{Álvaro Becerra, Roberto Daza, Ruth Cobos, Aythami Morales, Julian Fierrez}
\affiliation{%
  \institution{Universidad Autónoma de Madrid\country{España}}
  }
  
\email{{alvaro.becerra,roberto.daza, ruth.cobos, aythami.morales, julian.fierrez}@uam.es}

\renewcommand{\shortauthors}{Trovato et al.}


\begin{abstract}
En este artículo, presentamos M2LADS, un sistema que permite la integración y visualización de datos multimodales en forma de \textit{Dashboards} Web. Estos datos provienen de sesiones de experiencia de usuario en un sistema de \textit{Learning Analytics} (LA) llevadas a cabo por estudiantes de MOOCs. Los datos multimodales incluyen señales biométricas y de comportamiento monitorizados por la plataforma edBB, como electroencefalogramas (EEG) de 5 canales, frecuencia cardíaca, atención visual, videos en el espectro visible y NIR, entre otros. Además, se incluyen datos de interacción de los estudiantes con el sistema de LA a través de la herramienta LOGGE. Toda esta información proporciona una comprensión completa de la experiencia del usuario al utilizar el sistema de LA, lo que ha permitido tanto mejorar el sistema LA como la experiencia de aprendizaje de los estudiantes de MOOCs.
\end{abstract}




\maketitle

\section{Introducción}

Los \textit{Massive Open Online Courses} (MOOCs) son una valiosa fuente de contenido educativo y son respaldados y reconocidos cada vez más por instituciones oficiales~\cite{ma2019investigating}. 

Según múltiples estudios de investigación, uno de los principales problemas de los MOOCs es la falta de interacción entre estudiantes y profesores, además, los estudiantes admiten tener un sentimiento de soledad, falta de apoyo y falta de retroalimentación por parte de los profesores~\cite{pardo2019using,iraj2020understanding}. Esto hace que la motivación de los estudiantes disminuya, lo que finalmente lleva a que abandonen el curso~\cite{hone2016exploring,topali2019exploring}.

Un amplio número de instituciones españolas ofrecen MOOCs en diversas plataformas. En el caso de la Universidad Autónoma de Madrid (UAM), dichos cursos son ofrecidos a través de la plataforma edX\footnote{\url{https://www.edx.org/}}. Con el objetivo de minimizar los problemas antes mencionados, en la UAM hemos desarrollado un sistema de \textit{Learning Analytics} (LA)~\cite{lang2017handbook,martinez2020achievements,romeroeducational} para sus MOOCs con dos objetivos principales. 

En primer lugar, un objetivo del sistema desarrollado es proporcionar a los profesores de MOOCs la capacidad de realizar un seguimiento de sus alumnos y así poder intervenir en los procesos de aprendizaje de sus estudiantes. Por esta razón, el sistema se denomina edX-LIMS, que es la abreviatura de \textit{``System for Learning Interventions and Its Monitoring for edX MOOCs''} (``sistema para las intervenciones de aprendizaje y su monitorización en MOOCs en edX'')~\cite{cobos2020proposal,cobos2022learning,pascual2022proposal,cobos2021improving}. En segundo lugar, otro objetivo del sistema es proporcionar retroalimentación a los estudiantes de MOOCs para mejorar su motivación, persistencia y participación en el curso en el MOOC. edX-LIMS facilita tanto a profesores como a estudiantes un \textit{Dashboard} Web personalizado donde o bien realizar el seguimiento de sus estudiantes o bien ver su desempeño en los MOOCs en el caso de los estudiantes.

Este sistema de LA se ha utilizado durante aproximadamente tres años en un curso que está dirigido a aquellos interesados en aprender a desarrollar aplicaciones web. Aunque se ha comprobado la gran utilidad de edX-LIMS tanto para estudiantes como profesores del curso, estamos interesados en mejorar las visualizaciones de los \textit{dashboards}~\cite{matcha2019systematic,verbert2013learning,verbert2014learning} que genera, así como simplificar y clarificar la información que ofrece.

La captación de datos multimodales puede ser una excelente manera de obtener información valiosa sobre cómo los estudiantes interactúan con los \textit{Dashboards} y cómo se sienten al hacerlo. Con estos datos, podemos analizar patrones y tendencias que permitan simplificar y mejorar la información que se muestra gracias al Análisis de Aprendizaje Multimodal (MMLA, \textit{Multimodal Learning Analytics})~\cite{giannakos2022multimodal,giannakos2023role,spikol2018supervised}, lo que a su vez podría mejorar la experiencia del usuario y aumentar la eficacia del sistema. Además, puede identificar áreas que se necesitan mejorar y ajustar para ayudar a los estudiantes a obtener el máximo beneficio de la información mostrada en su \textit{Dashboard}.

Por ello, hemos realizado la captura de datos multimodales durante el uso de los \textit{Dashboards} ofrecidos por edX-LIMS por parte de una serie de estudiantes del MOOC. Para la integración de estos datos multimodales hemos desarrollado otro sistema llamado M2LADS, que es un acrónimo de \textit{``Multimodal Learning Analytics Dashboards System''} (``sistema para Generar \textit{Dashboards} de Análisis de Aprendizaje Multimodal''). Gracias a los \textit{Dashboards} que genera M2LADS, los profesores del MOOC pueden realizar un análisis exhaustivo para comprender mejor cuando están los estudiantes concentrados mientras navegan por los \textit{Dashboards} ofrecidos por edX-LIMS, y qué contenido capta su atención, entre otras cosas. En resumen, el sistema M2LADS complementa la información disponible en plataformas de \textit{Learning Analytics} como edX-LIMS y permite analizar factores relacionados con el estado cognitivo y emocional de los estudiantes. 

La estructura de este artículo es la siguiente: en la sección ~\ref{sec:edX-LIMS}, presentamos en detalle el sistema de LA edX-LIMS. En la sección ~\ref{sec:M2LADS} proporcionamos una descripción detallada del sistema propuesto M2LADS. En la sección~\ref{SEC:caso_estudio}, presentamos el caso de estudio y los resultados obtenidos de la captación de datos multimodales y el uso del sistema propuesto M2LADS. Finalmente, el artículo concluye con las conclusiones y trabajo futuro.
\section{edX-LIMS}\label{sec:edX-LIMS}
edX-LIMS es un sistema de \textit{Learning Analytics} (LA) cuyo nombre proviene del acrónimo de: \textit{``System for Learning Interventions and Its Monitoring for edX MOOCs''}~\cite{cobos2020proposal, pascual2022proposal,cobos2022learning,cobos2021improving}. Este sistema permite a los profesores de MOOCs monitorizar el progreso de sus estudiantes mediante un \textit{Dashboard} Web y llevar a cabo una estrategia de intervención en su aprendizaje. Además, edX-LIMS proporciona a los estudiantes del MOOC una amplia información sobre su interacción en el curso y cómo se espera que siga siendo en el futuro, todo ello reflejado en un \textit{Dashboard} Web.

\begin{figure}[t]
\centering
\includegraphics[width=6cm]{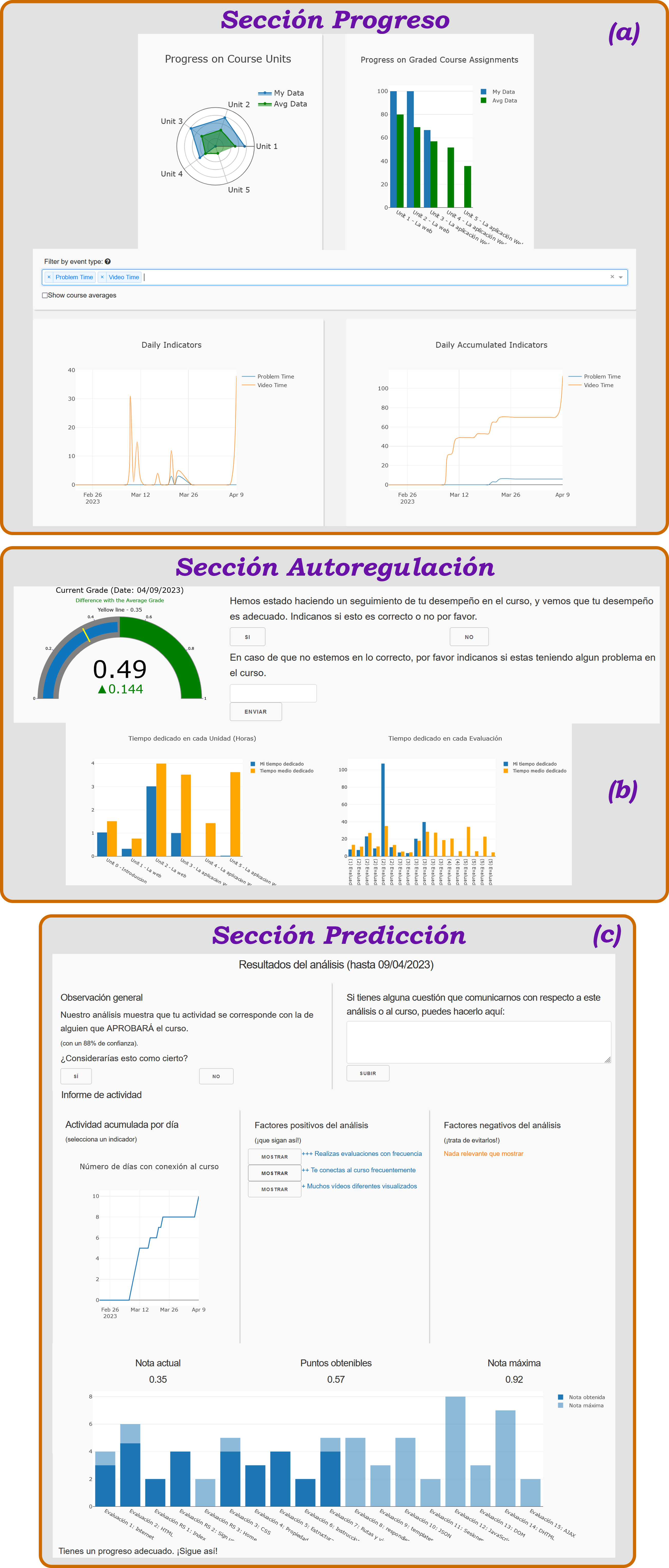} 
\caption{Ejemplos de gráficas mostradas en cada una de las secciones de un \textit{dashboard} de la plataforma edX-LIMS}
\label{dashboard_edx_lims}
\end{figure}

En el caso de los estudiantes, este sistema LA proporciona a cada uno un \textit{Dashboard} donde pueden ver lo siguiente en las secciones que le componen:

\begin{itemize}
    \item Progreso (Fig. \ref{dashboard_edx_lims} (a)): edX-LIMS proporciona al usuario información detallada sobre su progreso y desempeño en el curso, a través de gráficas que muestran cómo es su interacción con el mismo. Estas gráficas se generan a partir de una serie de valores (o indicadores) que se calculan a partir de los datos de registro (\textit{logs}) del MOOC. Por ejemplo, el sistema muestra el número de sesiones que el usuario ha tenido en el curso a lo largo del tiempo, así como el número de eventos que ha generado mientras visualizaba vídeos o resolvía ejercicios, entre otros datos relevantes~\cite{cobos2020proposal}.
    \item Autorregulación (Fig. \ref{dashboard_edx_lims} (b)): edX-LIMS detecta si el estudiante está teniendo algún problema de autorregulación en su proceso de aprendizaje y le informa de ello, ofreciéndole la posibilidad de proporcionar retroalimentación (\textit{feedback}) al profesor sobre la información recibida mediante unos botones de si/no (confirmando o no la información) y un cuadro de texto. Además, el sistema proporciona gráficas donde el estudiante puede ver el tiempo que ha dedicado y el número de veces que ha accedido a las diferentes secciones del curso, vídeos y ejercicios~\cite{cobos2022learning}.
    \item Predicción (Fig. \ref{dashboard_edx_lims} (c)): edX-LIMS proporciona información al estudiante sobre su probabilidad de aprobar el curso o si por el contrario se ha detectado que tiene dificultades para seguir el ritmo del curso y por tanto podría abandonarlo. Asimismo, el sistema le ofrece la oportunidad de proporcionar retroalimentación al profesor sobre esta predicción mediante unos botones de si/no  y un cuadro de texto. Además, el sistema muestra la progresión del estudiante en las calificaciones de los ejercicios del curso y le informa de la posible nota que podría llegar a alcanzar~\cite{pascual2022proposal}.
\end{itemize}

Tanto para la sección de Progreso como para la sección de Autorregulación, edX-LIMS utiliza analítica descriptiva, lo que significa que muestra visualmente lo que el estudiante ha realizado en el curso hasta ese momento a través de diversas gráficas. Además, el sistema le permite comparar sus valores con los de otros estudiantes del curso en la mayoría de estas gráficas, lo que le permite tener una mejor comprensión de su desempeño en relación con sus compañeros.

En la sección de Predicción, el sistema utiliza analítica predictiva y emplea el algoritmo \textit{Random Forest}~\cite{breiman2001random}, el cual ha sido entrenado con datos de más de dos años para desarrollar un modelo de predicción altamente preciso. Utilizando este modelo, el sistema puede predecir en tiempo real si un estudiante tiene posibilidades de aprobar el curso y, por lo tanto, obtener un certificado, o si está en riesgo de abandonarlo.

Tanto en la sección de Autorregulación como en la sección de Predicción, el estudiante puede dar \textit{feedback} sobre la información relacionada con la detección del problema y la predicción a través de dos vías: por un lado, puede dar su conformidad o no pulsando en unos botones que ofrece el \textit{Dashboard} para tal efecto, y por otro lado, puede escribir un texto relacionado con esta situación en un cuadro de texto que también ofrece el sistema. Gracias a este servicio, tanto los estudiantes como los profesores se sienten más conectados.

\section{M2LADS}\label{sec:M2LADS}
Proponemos un sistema web llamado M2LADS (acrónimo de \textit{``System for generating Multimodal Learning Analytics Dashboards''}). Este sistema permite generar \textit{Dashboards} Web enriquecidos con información de comportamiento de los usuarios. Cada \textit{Dashboard} visualiza todos los datos multimodales de un estudiante monitorizado en una sesión de experiencia de usuario (UX, \textit{User eXperience})~\cite{zarour2017user,Hassenzahl2006} en la plataforma edX-LIMS (ver Sección~\ref{SEC:caso_estudio} para más detalles).

El sistema se compone de tres módulos siguiendo el modelo Modelo-Vista-Controlador (MVC). Los datos multimodales capturados en la sesión y procesados por el sistema se han denominado Datos por Actividad (DA). Los detalles de los datos multimodales se encuentran en la Sección~\ref{sec:Descripción de Datos Multimodales}.

Los Módulos del sistema M2LADS son (Fig. \ref{sec:M2LADS}):

\begin{itemize}
 
\item Módulo de Procesamiento de Datos (Controlador): ver Sección~\ref{sec:Módulo de Procesamiento de AD}.

\item Módulo de Gestión de los Datos por Actividades (Modelo): ver Sección~\ref{sec:Módulo de gestión de AD}.

\item Módulo de Visualización de los Datos por Actividades (Vista): véase el apartado~\ref{sec:Módulo de Visualización AD}.
\end{itemize}

\subsection{Descripción de Datos Multimodales} \label{sec:Descripción de Datos Multimodales}

\subsubsection{Datos de edBB}
\begin{figure}[t]
 \centering
  \includegraphics[width=\linewidth]{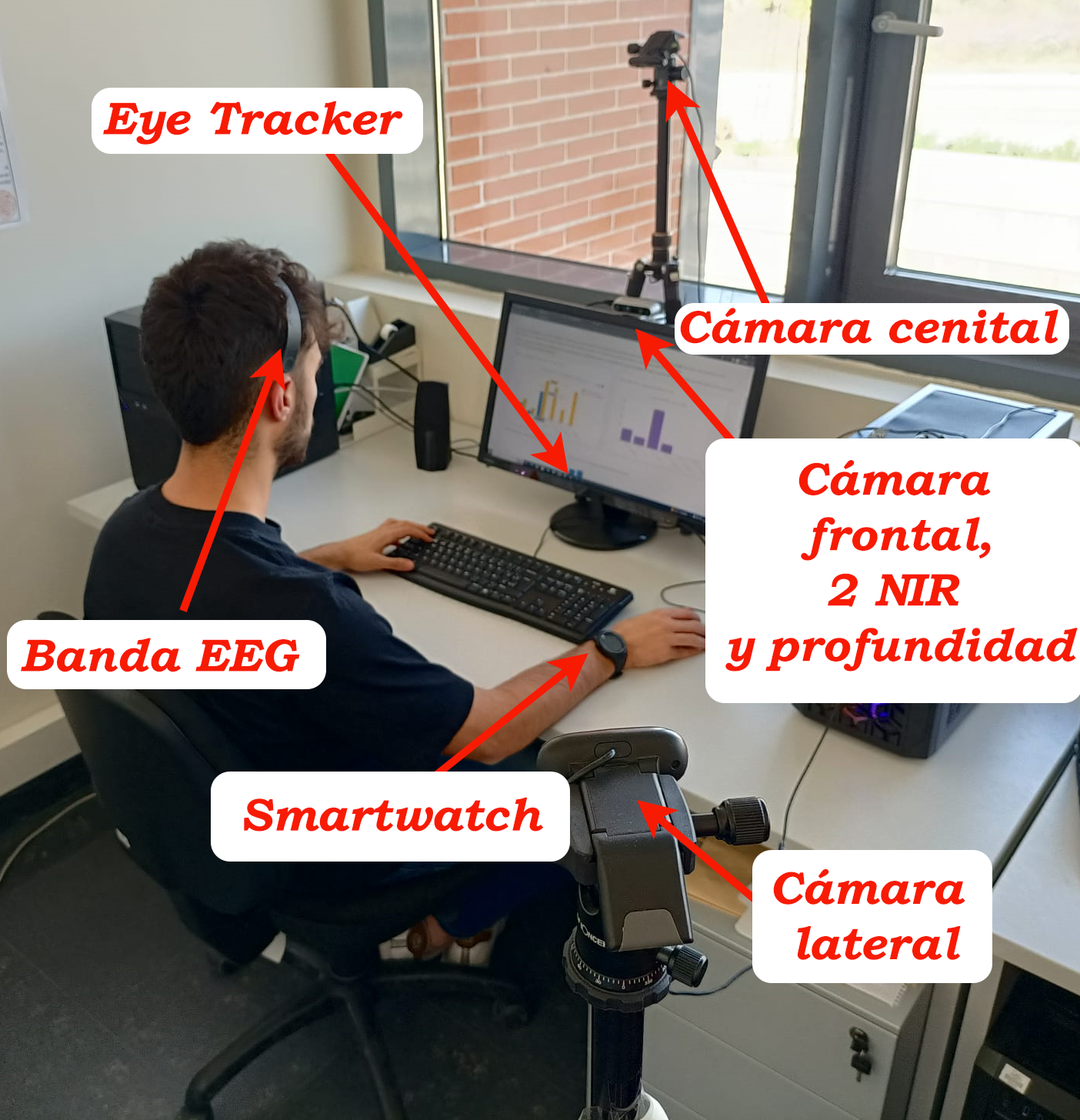} 
  \caption{Imágen con los sensores utilizados en una sesión de monitorización con la plataforma edBB~\cite{hernandez2019edbb}.}
   \label{Setup_edBB}
\end{figure}

Se ha utilizado la plataforma edBB para monitorizar a los estudiantes durante las actividades de la sesión de experiencia de usuario~\cite{daza2023edbb,hernandez2019edbb}. Este marco de adquisición multimodal fue diseñado para la monitorización de estudiantes en educación a distancia, capturando información biométrica y de comportamiento.
edBB está formado por un conjunto de programas que permiten comunicar y utilizar de forma sincronizada diferentes sensores y adaptar la configuración de adquisición a las circunstancias de monitorización, desde poder usar sensores avanzados (relojes inteligentes o \textit{smartwatches}, cámaras de seguimiento de mirada o \textit{eye-trackers}, etc.), básicos (cámaras web, datos de contexto, etc.) o ambos. En nuestro trabajo, hemos utilizado la siguiente configuración de adquisición y fuentes de información/sensores (ver Fig.~\ref{Setup_edBB}):
\begin{itemize}
\item Vídeo: Datos de vídeo procedentes de $3$ cámaras en diferentes posiciones: Cenital, frontal y lateral, utilizando $2$ cámaras web simples y $1$ Intel RealSense que incluye $1$ cámaras RGB (espectro visible) y $2$ cámaras NIR (\textit{Near Infrared}); además este dispositivo calcula las imágenes de profundidad utilizando las cámaras NIR. Por último, también se captura el vídeo de monitorización de la pantalla.
\item Datos del electroencefalograma (EEG): Utilizamos una  banda de EEG NeuroSky que obtiene señales $5$: $\delta$~($<4$Hz), $\theta$~($4$-$8$ Hz), $\alpha$~($8$-$13$ Hz), $\beta$~($13$-$30$ Hz), y $\gamma$~($>30$ Hz) y mediante el preprocesamiento de estos canales de EEG, la banda estima los niveles de atención y de meditación y el momento en que se producen los parpadeos~\cite{daza2022alebk,daza2023matt,daza2020mebal}. 
\item Ritmo cardíaco: Para obtener la frecuencia cardiaca en tiempo real utilizamos el pulsómetro de un Huawei Watch $2$~\cite{hernandez2020heart}.
\item Actividad del ratón y teclado: edBB registra los eventos del teclado y ratón (pulsar/soltar tecla, posición del ratón, etc.) que son un rasgo biométrico de comportamiento de gran valor \cite{morales2016kboc}. Igualmente son utilizados para detectar la actividad del estudiante. Además, también monitorizamos los movimientos que el usuario realiza con el ratón a través de los sensores inerciales del reloj inteligente (acelerómetro, giroscopio, etc.).
\item Atención visual: Se utilizó un Tobii Pro Fusion que contiene dos cámaras de seguimiento ocular que estiman los siguientes datos: Origen y punto de la mirada, diámetro de la pupila, calidad de los datos, etc.; permitiéndonos obtener la atención visual.
\item Metadatos: Los recogemos de sesiones y PCs como direcciones IP y MAC, SO, apps e historial web.
\end{itemize}

\begin{figure*}[t]
 \centering
\includegraphics[width=\linewidth]{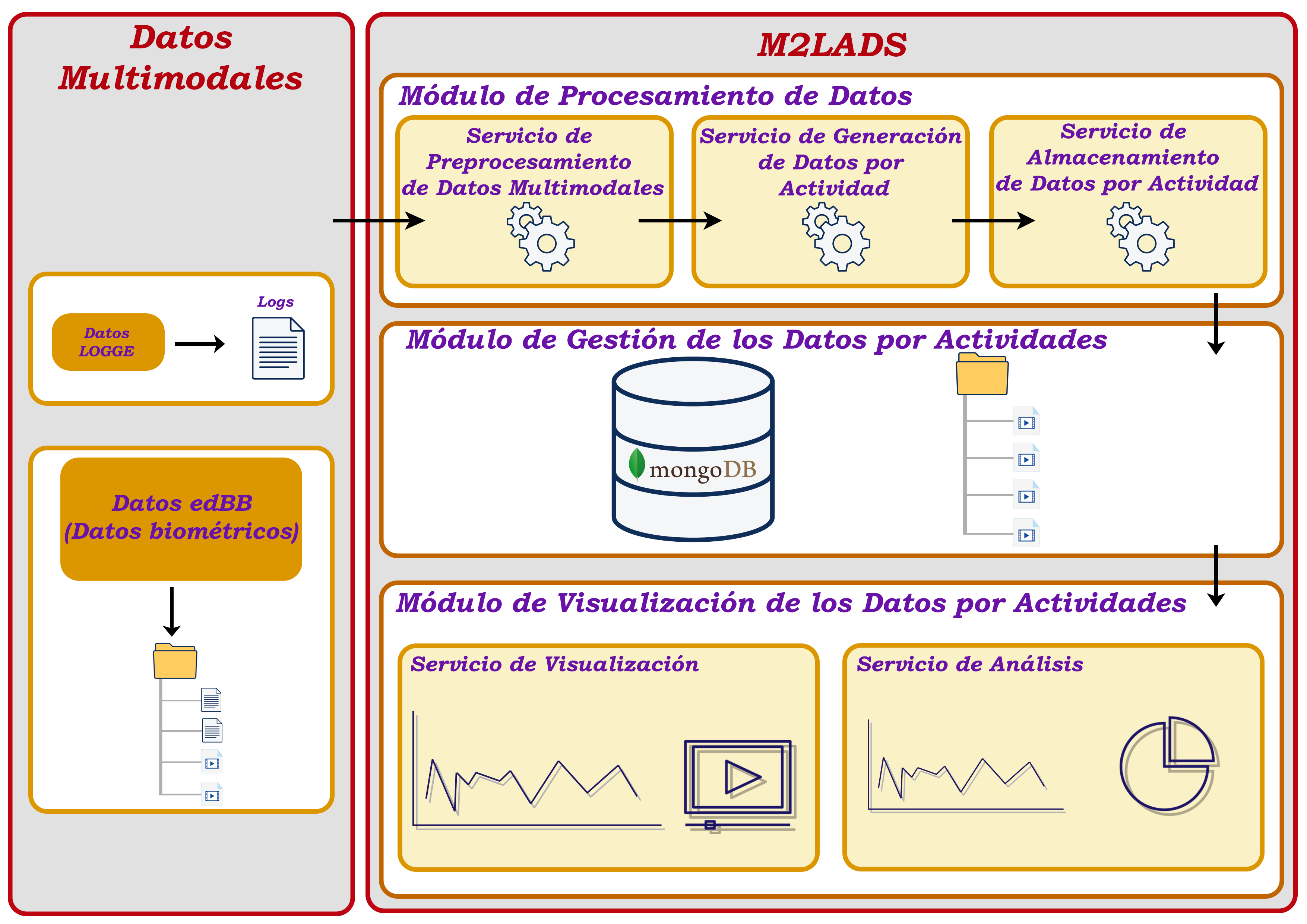} 
  \caption{Arquitectura y módulos del sistema M2LADS}
   \label{arquitectura_M2LAD}
\end{figure*}

\subsubsection{Datos adicionales. Herramienta LOGGE}
El objetivo de esta herramienta es almacenar información adicional relacionada con el estudiante monitorizado (por ejemplo, sexo, mano utilizada con el ratón, problemas cardíacos, grupo, etc.) y registrar las interacciones que realiza delante del ordenador durante la sesión UX.

\subsection{Módulo de Procesamiento de Datos} \label{sec:Módulo de Procesamiento de AD}

Este módulo permite extraer, limpiar, seleccionar y preprocesar los datos multimodales registrados durante la sesión UX para extraer los Datos por Actividad del estudiante. El módulo se compone de los tres servicios siguientes:

\subsubsection{Servicio de Preprocesamiento de los Datos Multimodales} 
Este servicio se encarga de extraer y procesar datos de dos fuentes:
\begin{itemize}
\item Datos LOGGE: El archivo de \textit{logs} generado por la herramienta LOGGE, que contiene información relacionada con las actividades o eventos que el estudiante ha realizado y los instantes de tiempo en el que han ocurrido, se preprocesa en una matriz de actividades (Matriz de Actividades) con tres columnas: el identificador de la actividad, el tiempo de inicio y el tiempo de fin de cada actividad.
\item Datos edBB: Las señales biológicas y de comportamiento proporcionadas por los dispositivos biométricos se preprocesan de la siguiente manera: para cada variable, como la atención (banda EEG), la frecuencia cardiaca (\textit{smartwatch}) o el diámetro de la pupila (\textit{eye-tracker}), se genera una matriz con estas columnas: instante de tiempo, valor de la variable y ventana de tiempo (la media de los datos más recientes de los últimos 30 segundos). En todas estas matrices, la unidad de tiempo es la misma, y tiempo de inicio y de fin están sincronizados para que los datos de todas las variables empiecen y terminen simultáneamente. Además, se calculan las correlaciones. Para cada vídeo, se extrae el número de \textit{frame} fotogramas y el instante en el que se tomó y se genera una matriz por cada vídeo con esta información en dos columnas.

\end{itemize}

\subsubsection{Servicio de Generación de Datos por Actividad} 
Este servicio cruza los datos biométricos de los estudiantes con la Matriz de Actividades para generar la Matriz del Estudiante (ME).
Para ello, cada una de las matrices con los datos de cada una de las variables (atención, frecuencia cardiaca, etc.) se combina con la matriz de actividad clasificando cada dato en la actividad que estaba realizando el estudiante en ese momento. La ME contiene cuatro columnas: instante de tiempo, valor de la variable, ventana de tiempo e identificador de la actividad.

\subsubsection{Servicio de Almacenamiento de Datos por Actividades}

Este servicio almacena la ME como colecciones de datos en una base de datos MongoDB, y todos los vídeos grabados durante la sesión se organizan como archivos audiovisuales en un conjunto de directorios.

\subsection{Módulo de Gestión de Datos por Actividades} \label{sec:Módulo de gestión de AD}
Este módulo proporciona conectividad con MongoDB y los directorios con los archivos audiovisuales.

\subsection{Módulo de Visualización de Datos por Actividades} \label{sec:Módulo de Visualización AD}

En este módulo, el sistema crea una visualización por cada estudiante, es decir, un \textit{Dashboard}, que refleja los datos por actividad del estudiante durante la sesión. Para ello, genera y organiza componentes visuales (gráficos) utilizando el \textit{framework} Dash\footnote{\url{https://plotly.com/dash/}}, basado en Flask\footnote{\url{https://flask.palletsprojects.com/en/2.3.x/}} y React.js\footnote{\url{https://es.react.dev/}}.
El módulo está compuesto por los siguientes servicios:

\begin{figure}[t]
 \centering
\includegraphics[width=\linewidth]{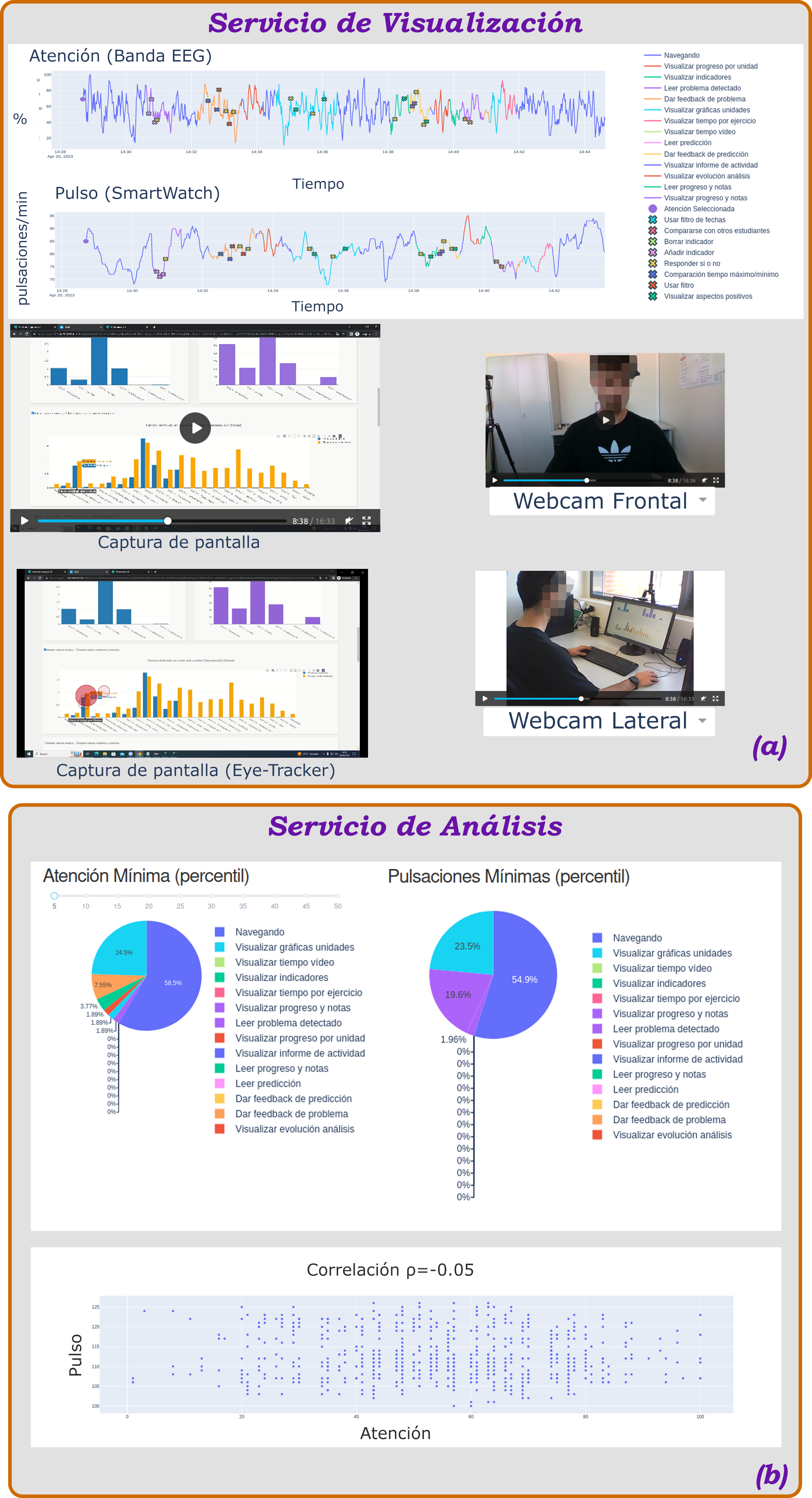} 
\caption{Algunas capturas de un \textit{Dashboard} generado por M2LADS}
\label{FRONT_END_M2LAD}
\end{figure}

\subsubsection{Servicio de Visualización}  

Como se muestra en la Fig. \ref{FRONT_END_M2LAD} (a), este servicio compone el \textit{dashboard} con varios gráficos que muestran las distintas variables biométricas como la atención del estudiante, la frecuencia cardíaca o el diámetro de la pupila. También incluye varios vídeos, como la pantalla capturada durante la sesión, distintos vídeos de las \textit{webcams} y el vídeo de las áreas de fijación del estudiante procedente del \textit{eye-tracker}. Todos estos elementos están sincronizados entre sí.

\subsubsection{Servicio de Análisis} 
Como se observa en la Fig. \ref{FRONT_END_M2LAD} (b) este servicio añade gráficos al \textit{dashboard} que muestran un análisis de los datos anteriores mediante percentiles y un análisis de la existencia o no de correlaciones entre los valores de las señales biométricas.

\section{Caso de estudio}\label{SEC:caso_estudio}

\begin{figure}[t]
 \centering
  \includegraphics[width=\linewidth]{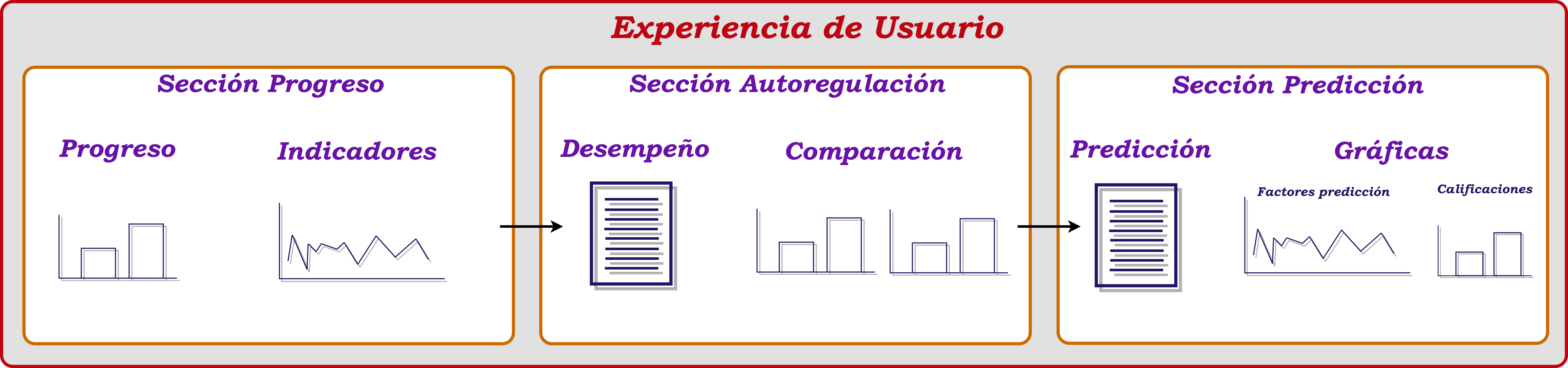} 
  \caption{Esquema de la sesión de la experiencia de usuario}
   \label{UX}
\end{figure}

El sistema presentado en este artículo ha sido probado con estudiantes el MOOC ''Introducción al desarrollo de aplicaciones web'' disponible en la plataforma edX y ofrecido por la UAM. Los estudiantes han sido monitorizados durante una sesión de UX con su \textit{dashboard} generado por el sistema de LA edX-LIMS (véase sección~\ref{sec:edX-LIMS}). Estos \textit{dashboards} están compuestos por tres secciones: progreso, autorregulación y predicción. El objetivo es monitorizar un total de 20 estudiantes (10 hombres y 10 mujeres).

Para ello, hemos diseñado la UX siguiendo el esquema que se muestra en la Fig.~\ref{UX}. Al inicio de la sesión, los estudiantes acceden a la sección de \textit{Progreso} para observar su progreso en el curso semana a semana y estudiar los valores de sus indicadores a lo largo del tiempo, con la posibilidad de configurar la gráfica para mostrar más o menos información (indicadores). Después, avanzan a la sección \textit{Autorregulación}, donde leen la información que facilita el sistema sobre la detección de posibles problemas en su autorregulación del aprendizaje y envían \textit{feedback} sobre dicha situación al profesor. Entonces, exploran las distintas gráficas y comparan su tiempo dedicado a cada apartado del curso con la media del resto de los estudiantes. Finalmente, acceden a la sección \textit{Predicción}, donde leen la predicción facilitada por el sistema, enviar su \textit{feedback} sobre la misma y exploran las distintas gráficas presentadas.

\begin{figure}[t]
 \centering
  \includegraphics[width=\linewidth]{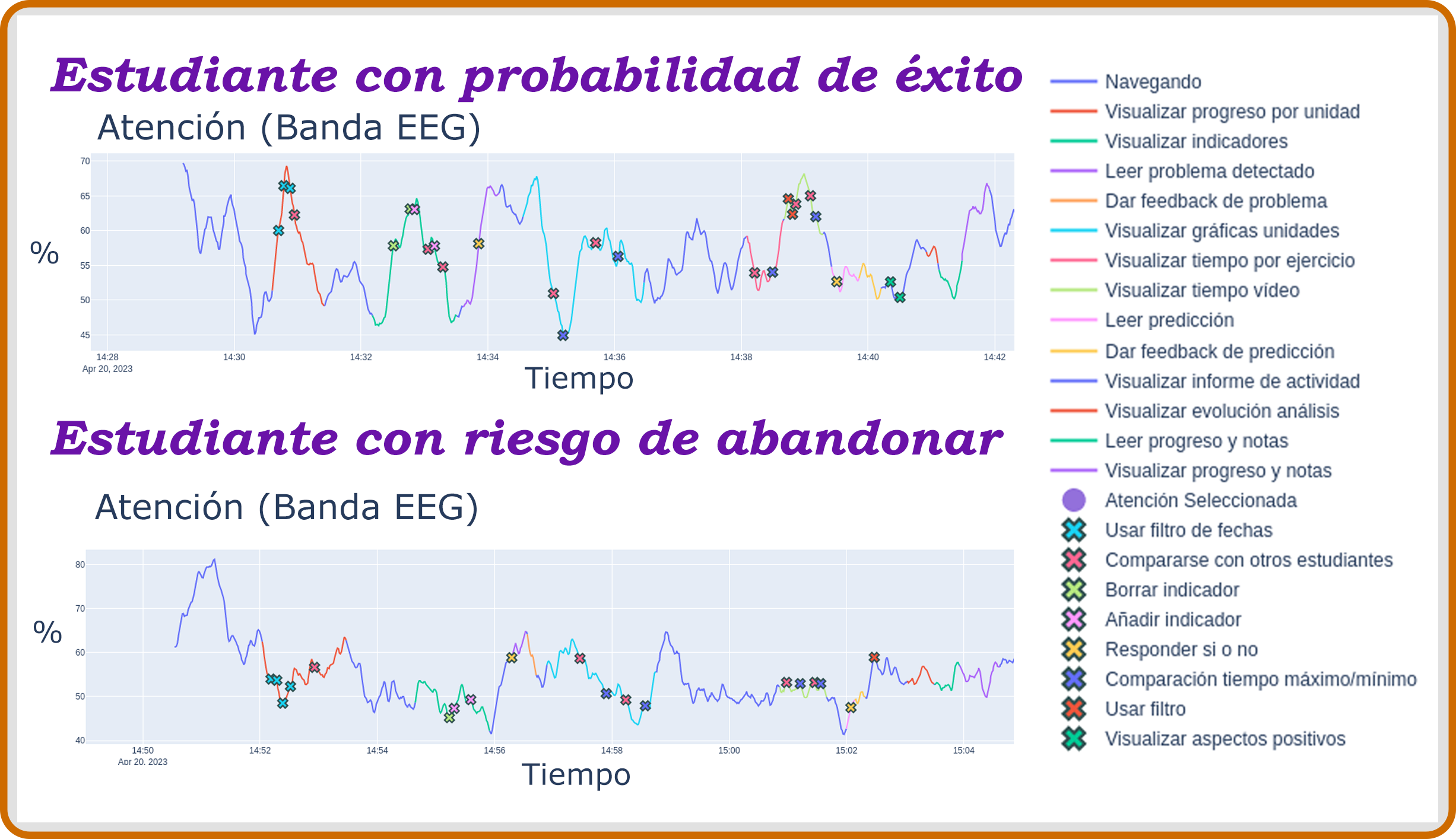} 
  \caption{Comparación del nivel de atención de dos estudiantes}
   \label{comparacion}
\end{figure}

Todos los datos capturados desde la plataforma edBB son procesados mediante el sistema M2LADS para generar gráficas y vídeos sincronizados, tal y como se muestra en el ejemplo de la Fig.~\ref{FRONT_END_M2LAD}. A partir de las monitorizaciones realizadas, hemos extraído de los \textit{dashboards} de M2LADS las gráficas que se presentan en la Fig. \ref{comparacion}. 

La gráfica superior de la Fig. \ref{comparacion} corresponde a un estudiante (estudiante 1) que ha demostrado un buen desempeño en el MOOC, con una nota cercana al aprobado y con una alta probabilidad de aprobar el curso según la predicción de edX-LIMS. Si el estudiante mantiene su ritmo actual, se espera que termine el curso con éxito. A los estudiantes con perfiles similares a este se les puede categorizar como ``estudiantes con probabilidad de éxito''. 

Por otro lado, la gráfica inferior de la Fig. \ref{comparacion} corresponde a otro estudiante (estudiante 2) que ha estado realizando el curso durante varios meses, pero no ha interactuado con él en el último mes. edX-LIMS ha pronosticado que existe la posibilidad de que el estudiante abandone el curso. A los estudiantes con perfiles similares a este, se les puede clasificar como ``estudiantes con riesgo de abandonar''.

Como se aprecia en la Fig. \ref{comparacion}, el estudiante 1 presenta unos niveles de atención superiores al estudiante 2. Además, el estudiante 1 muestra un mayor interés al interactuar con las gráficas del \textit{dashboard}, lo que provoca picos de atención, mientras que el estudiante 2 apenas muestra cambios en su nivel de interés. En particular, el estudiante 1 presta especial atención a las siguientes secciones del \textit{dashboard}:
\begin{itemize}
    \item Cuando visualiza su progreso por unidad y selecciona rangos de fechas para visualizar estos datos.
    \item Cuando visualiza sus indicadores del curso y, en concreto, cuando interactúa añadiendo nuevos indicadores.
    \item Cuando lee la información sobre la posible detección de problema.
    \item Cuando visualiza el tiempo dedicado en cada vídeo y se compara con otros estudiantes.
    \item Cuando visualiza sus calificaciones del curso.
\end{itemize}

Por otro lado, el estudiante 2 solo muestra una atención significativa al leer y escribir su \textit{feedback} sobre la detección de problema.

Tras la sesión de UX, se solicita a los estudiantes que completen una encuesta de satisfacción con el fin de valorar las diversas secciones y gráficas presentes en el \textit{dashboard} creado por edX-LIMS. De acuerdo con el análisis de las respuestas obtenidas, de manera preliminar podemos destacar los siguientes aspectos:
\begin{itemize}
    \item La mayoría de los estudiantes valoran positivamente el acceso a edX-LIMS para visualizar información sobre su progreso en el curso. También afirman sentirse más orientados y supervisados, lo que reduce su sensación de soledad durante el proceso de aprendizaje.
    \item Respecto a la influencia del \textit{dashboard} en su motivación, los estudiantes con mayores probabilidades de éxito afirman sentirse motivados por la plataforma. Sin embargo, aquellos con mayor riesgo de abandono no sienten que el \textit{dashboard} les anime a participar más activamente en el curso o a trabajar de manera más constante.   
    \item Sobre las predicciones y su nivel de desempeño, los estudiantes con mayores probabilidades de éxito consideran que el sistema acierta, mientras que aquellos con mayor riesgo de abandono no están tan convencidos.
\end{itemize}

\section{Conclusión y Trabajo Futuro}\label{sec:Conclusión y Trabajo Futuro}

En este artículo, presentamos un sistema web llamado M2LADS (acrónimo de \textit{System for Generating Multimodal Learning Analytics Dashboards}). Este sistema soporta la integración y visualización de datos multimodales en forma de \textit{Dasboards} o cuadros de mando resultantes de la medición y análisis de la experiencia de usuario en plataformas de \textit{Learning Analytics} por parte de estudiantes de un MOOC. Estos datos multimodales son proporcionados por la monitorización soportada por la plataforma edBB que facilita información biométrica y de comportamiento y los datos de la herramienta LOGGE que registra las interacciones de los usuarios con el sistema durante dichas monitorizaciones. 

En conclusión, en este trabajo se presenta un laboratorio MMLA completamente equipado que integra M2LADS, la herramienta LOGGE y la plataforma edBB. Hemos utilizado este laboratorio para llevar a cabo pruebas de usuario en el contexto de la interacción de los estudiantes con una herramienta de \textit{Learning Analytics} que proporciona información sobre su progreso y desempeño en un MOOC. Este enfoque ha permitido monitorizar y mejorar la experiencia de aprendizaje de los estudiantes, además de brindar valiosos datos para futuras investigaciones en el área de \textit{Learning Analytics}. 

El sistema desarrollado permitirá incorporar nuevos parámetros a las plataformas de \textit{Learning Analytics} y avanzar el conocimiento para el uso de herramientas de \textit{Learning Analytics} en el contexto de los MOOCs. Líneas de trabajo futuro incluyen el estudio de la efectividad de estos sistemas para mejorar el proceso de enseñanza-aprendizaje y mejorar la experiencia de usuario en el ámbito educativo.

\section{Agradecimientos}

Trabajo financiado por los proyectos: HumanCAIC (TED2021-131787B-I00 MICINN), BIO-PROCTORING (GNOSS Program, Agreement Ministerio de Defensa-UAM-FUAM dated 29-03-2022), IndiGo! (PID2019-105951RB-I00), SNOLA (RED2022-134284-T) y e-Madrid-CM (S2018/TCS-4307, cofinanciada con Fondos Estructurales Europeos, FSE and FEDER). Roberto Daza ha sido financiado por una beca FPI, MINECO/FEDER.

\bibliographystyle{ACM-Reference-Format}

\bibliography{sample-base}


\begin{thebibliography}{28}


\ifx \showCODEN    \undefined \def \showCODEN     #1{\unskip}     \fi
\ifx \showDOI      \undefined \def \showDOI       #1{#1}\fi
\ifx \showISBNx    \undefined \def \showISBNx     #1{\unskip}     \fi
\ifx \showISBNxiii \undefined \def \showISBNxiii  #1{\unskip}     \fi
\ifx \showISSN     \undefined \def \showISSN      #1{\unskip}     \fi
\ifx \showLCCN     \undefined \def \showLCCN      #1{\unskip}     \fi
\ifx \shownote     \undefined \def \shownote      #1{#1}          \fi
\ifx \showarticletitle \undefined \def \showarticletitle #1{#1}   \fi
\ifx \showURL      \undefined \def \showURL       {\relax}        \fi
\providecommand\bibfield[2]{#2}
\providecommand\bibinfo[2]{#2}
\providecommand\natexlab[1]{#1}
\providecommand\showeprint[2][]{arXiv:#2}

\bibitem[Breiman(2001)]%
        {breiman2001random}
\bibfield{author}{\bibinfo{person}{Leo Breiman}.}
  \bibinfo{year}{2001}\natexlab{}.
\newblock \showarticletitle{Random forests}.
\newblock \bibinfo{journal}{\emph{Machine Learning}} \bibinfo{volume}{45},
  \bibinfo{number}{1} (\bibinfo{year}{2001}), \bibinfo{pages}{5--32}.
\newblock
\urldef\tempurl%
\url{https://doi.org/10.1023/A:1010933404324}
\showDOI{\tempurl}


\bibitem[Cobos(2022)]%
        {cobos2022learning}
\bibfield{author}{\bibinfo{person}{Ruth Cobos}.}
  \bibinfo{year}{2022}\natexlab{}.
\newblock \showarticletitle{The Learning Analytics System that Improves the
  Teaching-Learning Experience of MOOC Instructors and Students}. In
  \bibinfo{booktitle}{\emph{Proceedings of International Conference On
  Web-Based Learning - ICWL 2022}} \emph{(\bibinfo{series}{LNCS},
  Vol.~\bibinfo{volume}{13869})}. \bibinfo{address}{Tenerife, Spain},
  \bibinfo{pages}{29--40}.
\newblock
\urldef\tempurl%
\url{https://link.springer.com/chapter/10.1007/978-3-031-33023-0_3}
\showURL{%
\tempurl}


\bibitem[Cobos and Ruiz-Garcia(2021)]%
        {cobos2021improving}
\bibfield{author}{\bibinfo{person}{Ruth Cobos} {and} \bibinfo{person}{J.~C.
  Ruiz-Garcia}.} \bibinfo{year}{2021}\natexlab{}.
\newblock \showarticletitle{{Improving Learner Engagement in MOOCs using a
  Learning Intervention System: A Research Study in Engineering Education}}.
\newblock \bibinfo{journal}{\emph{Computer Applications in Engineering
  Education}} \bibinfo{volume}{29}, \bibinfo{number}{4} (\bibinfo{year}{2021}),
  \bibinfo{pages}{733--749}.
\newblock


\bibitem[Cobos and Soberón(2020)]%
        {cobos2020proposal}
\bibfield{author}{\bibinfo{person}{Ruth Cobos} {and} \bibinfo{person}{Juan
  Soberón}.} \bibinfo{year}{2020}\natexlab{}.
\newblock \showarticletitle{A Proposal for Monitoring the Intervention Strategy
  on the Learning of MOOC Learners}. In \bibinfo{booktitle}{\emph{CEUR
  Conference Proceedings}} \emph{(\bibinfo{series}{LASI-Spain 2020. Learning
  Analytics Summer Institute Spain 2020: Learning Analytics. Time for
  Adoption?}, Vol.~\bibinfo{volume}{2671})}. \bibinfo{address}{Valladolid,
  Spain}.
\newblock
\urldef\tempurl%
\url{http://ceur-ws.org/Vol-2671/paper07.pdf}
\showURL{%
\tempurl}


\bibitem[Daza et~al\mbox{.}(2022)]%
        {daza2022alebk}
\bibfield{author}{\bibinfo{person}{Roberto Daza}, \bibinfo{person}{Daniel
  DeAlcala}, \bibinfo{person}{Aythami Morales}, \bibinfo{person}{Ruben
  Tolosana}, \bibinfo{person}{Ruth Cobos}, {and} \bibinfo{person}{Julian
  Fierrez}.} \bibinfo{year}{2022}\natexlab{}.
\newblock \showarticletitle{{ALEBk: Feasibility Study of Attention Level
  Estimation Via Blink Detection Applied to e-learning}}. In
  \bibinfo{booktitle}{\emph{Proc. AAAI Workshop on Artificial Intelligence for
  Education}}.
\newblock


\bibitem[Daza et~al\mbox{.}(2023a)]%
        {daza2023matt}
\bibfield{author}{\bibinfo{person}{Roberto Daza}, \bibinfo{person}{Luis~F
  Gomez}, \bibinfo{person}{Aythami Morales}, \bibinfo{person}{Julian Fierrez},
  \bibinfo{person}{Ruben Tolosana}, \bibinfo{person}{Ruth Cobos}, {and}
  \bibinfo{person}{Javier Ortega-Garcia}.} \bibinfo{year}{2023}\natexlab{a}.
\newblock \showarticletitle{{MATT: Multimodal Attention Level Estimation for
  e-learning Platforms}}. In \bibinfo{booktitle}{\emph{Proc. AAAI Workshop on
  Artificial Intelligence for Education}}.
\newblock


\bibitem[Daza et~al\mbox{.}(2020)]%
        {daza2020mebal}
\bibfield{author}{\bibinfo{person}{Roberto Daza}, \bibinfo{person}{Aythami
  Morales}, \bibinfo{person}{Julian Fierrez}, {and} \bibinfo{person}{Ruben
  Tolosana}.} \bibinfo{year}{2020}\natexlab{}.
\newblock \showarticletitle{{mEBAL: A Multimodal Database for Eye Blink
  Detection and Attention Level Estimation}}. In
  \bibinfo{booktitle}{\emph{Proc. Intl. Conf. on Multimodal Interaction}}.
  \bibinfo{pages}{32--36}.
\newblock


\bibitem[Daza et~al\mbox{.}(2023b)]%
        {daza2023edbb}
\bibfield{author}{\bibinfo{person}{Roberto Daza}, \bibinfo{person}{Aythami
  Morales}, \bibinfo{person}{Ruben Tolosana}, \bibinfo{person}{Luis~F Gomez},
  \bibinfo{person}{Julian Fierrez}, {and} \bibinfo{person}{Javier
  Ortega-Garcia}.} \bibinfo{year}{2023}\natexlab{b}.
\newblock \showarticletitle{{edBB-Demo: Biometrics and Behavior Analysis for
  Online Educational Platforms}}. In \bibinfo{booktitle}{\emph{Proc. AAAI
  Conference on Artificial Intelligence (Demonstration)}}.
\newblock


\bibitem[Giannakos and Cukurova(2023)]%
        {giannakos2023role}
\bibfield{author}{\bibinfo{person}{Michail Giannakos} {and}
  \bibinfo{person}{Mutlu Cukurova}.} \bibinfo{year}{2023}\natexlab{}.
\newblock \showarticletitle{The role of learning theory in multimodal learning
  analytics}.
\newblock \bibinfo{journal}{\emph{British Journal of Educational Technology}}
  \bibinfo{volume}{00} (\bibinfo{year}{2023}), \bibinfo{pages}{1--22}.
\newblock
\urldef\tempurl%
\url{https://doi.org/10.1111/bjet.13320}
\showDOI{\tempurl}


\bibitem[Giannakos et~al\mbox{.}(2022)]%
        {giannakos2022multimodal}
\bibfield{editor}{\bibinfo{person}{Michail Giannakos}, \bibinfo{person}{Daniel
  Spikol}, \bibinfo{person}{Daniele Di~Mitri}, \bibinfo{person}{Kshitij
  Sharma}, \bibinfo{person}{Xavier Ochoa}, {and} \bibinfo{person}{Rawad
  Hammad}} (Eds.). \bibinfo{year}{2022}\natexlab{}.
\newblock \bibinfo{booktitle}{\emph{The Multimodal Learning Analytics
  Handbook}}.
\newblock \bibinfo{publisher}{Springer Nature}.
\newblock


\bibitem[Hassenzahl and Tractinsky(2006)]%
        {Hassenzahl2006}
\bibfield{author}{\bibinfo{person}{Marc Hassenzahl} {and} \bibinfo{person}{Noam
  Tractinsky}.} \bibinfo{year}{2006}\natexlab{}.
\newblock \showarticletitle{User experience - a research agenda}.
\newblock \bibinfo{journal}{\emph{Behaviour \& Information Technology}}
  \bibinfo{volume}{25}, \bibinfo{number}{2} (\bibinfo{year}{2006}),
  \bibinfo{pages}{91--97}.
\newblock
\urldef\tempurl%
\url{https://doi.org/10.1080/01449290500330331}
\showDOI{\tempurl}


\bibitem[Hernandez-Ortega et~al\mbox{.}(2020a)]%
        {hernandez2019edbb}
\bibfield{author}{\bibinfo{person}{Javier Hernandez-Ortega},
  \bibinfo{person}{Roberto Daza}, \bibinfo{person}{Aythami Morales},
  \bibinfo{person}{Julian Fierrez}, {and} \bibinfo{person}{Javier
  Ortega-Garcia}.} \bibinfo{year}{2020}\natexlab{a}.
\newblock \showarticletitle{{edBB: Biometrics and Behavior for Assessing Remote
  Education}}. In \bibinfo{booktitle}{\emph{Proc. AAAI Workshop on Artificial
  Intelligence for Education}}.
\newblock


\bibitem[Hernandez-Ortega et~al\mbox{.}(2020b)]%
        {hernandez2020heart}
\bibfield{author}{\bibinfo{person}{Javier Hernandez-Ortega},
  \bibinfo{person}{Roberto Daza}, \bibinfo{person}{Aythami Morales},
  \bibinfo{person}{Julian Fierrez}, {and} \bibinfo{person}{Ruben Tolosana}.}
  \bibinfo{year}{2020}\natexlab{b}.
\newblock \showarticletitle{{Heart Rate Estimation from Face Videos for Student
  Assessment: Experiments on edBB}}. In \bibinfo{booktitle}{\emph{Proc. Annual
  Computers, Software, and Applications Conference}}.
  \bibinfo{pages}{172--177}.
\newblock


\bibitem[Hone and El~Said(2016)]%
        {hone2016exploring}
\bibfield{author}{\bibinfo{person}{K.~S. Hone} {and} \bibinfo{person}{G.~R.
  El~Said}.} \bibinfo{year}{2016}\natexlab{}.
\newblock \showarticletitle{Exploring the factors affecting MOOC retention: A
  survey study}.
\newblock \bibinfo{journal}{\emph{Computers \& Education}}
  \bibinfo{volume}{98} (\bibinfo{year}{2016}), \bibinfo{pages}{157--168}.
\newblock
\urldef\tempurl%
\url{https://doi.org/10.1016/j.compedu.2016.03.016}
\showDOI{\tempurl}


\bibitem[Iraj et~al\mbox{.}(2020)]%
        {iraj2020understanding}
\bibfield{author}{\bibinfo{person}{Habib Iraj}, \bibinfo{person}{Amy Fudge},
  \bibinfo{person}{Michael Faulkner}, \bibinfo{person}{Abelardo Pardo}, {and}
  \bibinfo{person}{Vitomir Kovanovi{\'c}}.} \bibinfo{year}{2020}\natexlab{}.
\newblock \showarticletitle{Understanding Students’ Engagement with
  Personalised Feedback Messages}. In \bibinfo{booktitle}{\emph{LAK 2020
  Proceedings}}. ACM, \bibinfo{pages}{438--447}.
\newblock
\urldef\tempurl%
\url{https://doi.org/10.1145/3375462.3375527}
\showDOI{\tempurl}


\bibitem[Lang et~al\mbox{.}(2017)]%
        {lang2017handbook}
\bibfield{editor}{\bibinfo{person}{Charles Lang}, \bibinfo{person}{George
  Siemens}, \bibinfo{person}{Alyssa Wise}, {and} \bibinfo{person}{Dragan
  Gasevic}} (Eds.). \bibinfo{year}{2017}\natexlab{}.
\newblock \bibinfo{booktitle}{\emph{Handbook of Learning Analytics}}.
\newblock \bibinfo{publisher}{SOLAR, Society for Learning Analytics and
  Research New York}.
\newblock


\bibitem[Ma and Lee(2019)]%
        {ma2019investigating}
\bibfield{author}{\bibinfo{person}{Long Ma} {and} \bibinfo{person}{Chei~Sian
  Lee}.} \bibinfo{year}{2019}\natexlab{}.
\newblock \showarticletitle{{Investigating The Adoption of MOOCs: A
  Technology--User--Environment Perspective}}.
\newblock \bibinfo{journal}{\emph{Journal of Computer Assisted Learning}}
  \bibinfo{volume}{35}, \bibinfo{number}{1} (\bibinfo{year}{2019}),
  \bibinfo{pages}{89--98}.
\newblock


\bibitem[Mart{\'\i}nez~Mon{\'e}s et~al\mbox{.}(2020)]%
        {martinez2020achievements}
\bibfield{author}{\bibinfo{person}{Alejandra Mart{\'\i}nez~Mon{\'e}s},
  \bibinfo{person}{Ioannis Dimitriadis~Damoulis}, \bibinfo{person}{Emiliano
  Acquila~Natale}, \bibinfo{person}{Ainhoa {\'A}lvarez},
  \bibinfo{person}{Manuel Caeiro~Rodr{\'\i}guez}, \bibinfo{person}{Ruth
  Cobos~P{\'e}rez}, \bibinfo{person}{Miguel~{\'A}ngel Conde~Gonz{\'a}lez},
  \bibinfo{person}{Francisco~Jos{\'e} Garc{\'\i}a~Pe{\~n}alvo},
  \bibinfo{person}{Davinia Hern{\'a}ndez~Leo}, \bibinfo{person}{Iratxe
  Menchaca~Sierra}, {et~al\mbox{.}}} \bibinfo{year}{2020}\natexlab{}.
\newblock \showarticletitle{{Achievements and Challenges in Learning Analytics
  in Spain: The View of SNOLA}}.
\newblock \bibinfo{journal}{\emph{RIED. Revista Iberoamericana de Educaci{\'o}n
  a Distancia}} \bibinfo{volume}{23}, \bibinfo{number}{2}
  (\bibinfo{year}{2020}), \bibinfo{pages}{187}.
\newblock


\bibitem[Matcha et~al\mbox{.}(2019)]%
        {matcha2019systematic}
\bibfield{author}{\bibinfo{person}{Wannisa Matcha}, \bibinfo{person}{Dragan
  Ga{\v{s}}evi{\'c}}, \bibinfo{person}{Abelardo Pardo}, {et~al\mbox{.}}}
  \bibinfo{year}{2019}\natexlab{}.
\newblock \showarticletitle{A systematic review of empirical studies on
  learning analytics dashboards: A self-regulated learning perspective}.
\newblock \bibinfo{journal}{\emph{IEEE Transactions on Learning Technologies}}
  \bibinfo{volume}{13}, \bibinfo{number}{2} (\bibinfo{year}{2019}),
  \bibinfo{pages}{226--245}.
\newblock


\bibitem[Morales et~al\mbox{.}(2016)]%
        {morales2016kboc}
\bibfield{author}{\bibinfo{person}{Aythami Morales}, \bibinfo{person}{Julian
  Fierrez}, \bibinfo{person}{Marta Gomez-Barrero}, \bibinfo{person}{Javier
  Ortega-Garcia}, \bibinfo{person}{Roberto Daza}, \bibinfo{person}{John~V
  Monaco}, \bibinfo{person}{Jugurta Montalv{\~a}o}, \bibinfo{person}{J{\^a}nio
  Canuto}, {and} \bibinfo{person}{Anjith George}.}
  \bibinfo{year}{2016}\natexlab{}.
\newblock \showarticletitle{KBOC: Keystroke Biometrics Ongoing Competition}. In
  \bibinfo{booktitle}{\emph{Proceedings of the tenth international conference
  on learning analytics \& knowledge}}. \bibinfo{pages}{1--6}.
\newblock


\bibitem[Pardo et~al\mbox{.}(2019)]%
        {pardo2019using}
\bibfield{author}{\bibinfo{person}{Abelardo Pardo}, \bibinfo{person}{Jelena
  Jovanovic}, \bibinfo{person}{Shane Dawson}, \bibinfo{person}{Dragan
  Ga{\v{s}}evi{\'c}}, {and} \bibinfo{person}{Negin Mirriahi}.}
  \bibinfo{year}{2019}\natexlab{}.
\newblock \showarticletitle{Using learning analytics to scale the provision of
  personalised feedback}.
\newblock \bibinfo{journal}{\emph{British Journal of Educational Technology}}
  \bibinfo{volume}{50}, \bibinfo{number}{1} (\bibinfo{year}{2019}),
  \bibinfo{pages}{128--138}.
\newblock
\urldef\tempurl%
\url{https://doi.org/10.1111/bjet.12592}
\showDOI{\tempurl}


\bibitem[Pascual and Cobos(2022)]%
        {pascual2022proposal}
\bibfield{author}{\bibinfo{person}{Ivan. Pascual} {and} \bibinfo{person}{Ruth
  Cobos}.} \bibinfo{year}{2022}\natexlab{}.
\newblock \showarticletitle{A Proposal for Predicting and Intervening on MOOC
  Learners' Performance in Real Time}. In \bibinfo{booktitle}{\emph{CEUR
  Conference Proceedings}} \emph{(\bibinfo{series}{LASI-Spain 2022},
  Vol.~\bibinfo{volume}{3238})},
  \bibfield{editor}{\bibinfo{person}{A.~Vázquez-Ingelmo},
  \bibinfo{person}{Y.~Dimitriadis}, \bibinfo{person}{A.~Martínez-Monés},
  {and} \bibinfo{person}{F.~J. García-Peñalvo}} (Eds.).
  \bibinfo{address}{Salamanca, Spain}, \bibinfo{pages}{26--38}.
\newblock
\urldef\tempurl%
\url{https://ceur-ws.org/Vol-3238/paper4.pdf}
\showURL{%
\tempurl}


\bibitem[Romero and Ventura(2020)]%
        {romeroeducational}
\bibfield{author}{\bibinfo{person}{Cristóbal Romero} {and}
  \bibinfo{person}{Sebastian Ventura}.} \bibinfo{year}{2020}\natexlab{}.
\newblock \showarticletitle{{Educational Data Mining and Learning Analytics: An
  Updated Survey}}.
\newblock \bibinfo{journal}{\emph{Wiley Interdisciplinary Reviews: Data Mining
  and Knowledge Discovery}} (\bibinfo{year}{2020}).
\newblock


\bibitem[Spikol et~al\mbox{.}(2018)]%
        {spikol2018supervised}
\bibfield{author}{\bibinfo{person}{Daniel Spikol}, \bibinfo{person}{Emanuele
  Ruffaldi}, \bibinfo{person}{Giacomo Dabisias}, {and} \bibinfo{person}{Mutlu
  Cukurova}.} \bibinfo{year}{2018}\natexlab{}.
\newblock \showarticletitle{Supervised machine learning in multimodal learning
  analytics for estimating success in project-based learning}.
\newblock \bibinfo{journal}{\emph{Journal of Computer Assisted Learning}}
  \bibinfo{volume}{34}, \bibinfo{number}{4} (\bibinfo{year}{2018}),
  \bibinfo{pages}{366--377}.
\newblock


\bibitem[Topali et~al\mbox{.}(2019)]%
        {topali2019exploring}
\bibfield{author}{\bibinfo{person}{Paraskevi Topali},
  \bibinfo{person}{Alejandro Ortega-Arranz}, \bibinfo{person}{Erkan Er},
  \bibinfo{person}{Alejandra Mart{\'\i}nez-Mon{\'e}s}, \bibinfo{person}{Sara~L
  Villagr{\'a}-Sobrino}, {and} \bibinfo{person}{Yannis Dimitriadis}.}
  \bibinfo{year}{2019}\natexlab{}.
\newblock \showarticletitle{Exploring the problems experienced by learners in a
  MOOC implementing active learning pedagogies}. In
  \bibinfo{booktitle}{\emph{International Conference on Learning and
  Collaboration Technologies}} \emph{(\bibinfo{series}{LNCS},
  Vol.~\bibinfo{volume}{11475})}. Springer, \bibinfo{pages}{81--90}.
\newblock
\urldef\tempurl%
\url{https://doi.org/10.1007/978-3-030-19875-6_10}
\showDOI{\tempurl}


\bibitem[Verbert et~al\mbox{.}(2013)]%
        {verbert2013learning}
\bibfield{author}{\bibinfo{person}{Katrien Verbert}, \bibinfo{person}{Erik
  Duval}, \bibinfo{person}{Joris Klerkx}, \bibinfo{person}{Sten Govaerts},
  {and} \bibinfo{person}{Jos{\'e}~Luis Santos}.}
  \bibinfo{year}{2013}\natexlab{}.
\newblock \showarticletitle{Learning analytics dashboard applications}.
\newblock \bibinfo{journal}{\emph{American Behavioral Scientist}}
  \bibinfo{volume}{57}, \bibinfo{number}{10} (\bibinfo{year}{2013}),
  \bibinfo{pages}{1500--1509}.
\newblock


\bibitem[Verbert et~al\mbox{.}(2014)]%
        {verbert2014learning}
\bibfield{author}{\bibinfo{person}{Katrien Verbert}, \bibinfo{person}{Sten
  Govaerts}, \bibinfo{person}{Erik Duval}, \bibinfo{person}{Jose~Luis Santos},
  \bibinfo{person}{Frans Van~Assche}, \bibinfo{person}{Gonzalo Parra}, {and}
  \bibinfo{person}{Joris Klerkx}.} \bibinfo{year}{2014}\natexlab{}.
\newblock \showarticletitle{Learning dashboards: an overview and future
  research opportunities}.
\newblock \bibinfo{journal}{\emph{Personal and Ubiquitous Computing}}
  \bibinfo{volume}{18} (\bibinfo{year}{2014}), \bibinfo{pages}{1499--1514}.
\newblock


\bibitem[Zarour and Alharbi(2017)]%
        {zarour2017user}
\bibfield{author}{\bibinfo{person}{Mohammad Zarour} {and}
  \bibinfo{person}{Mubarak Alharbi}.} \bibinfo{year}{2017}\natexlab{}.
\newblock \showarticletitle{User Experience Aspects and Dimensions: Systematic
  Literature Review}.
\newblock \bibinfo{journal}{\emph{International Journal of Knowledge
  Engineering}} \bibinfo{volume}{3}, \bibinfo{number}{2}
  (\bibinfo{year}{2017}), \bibinfo{pages}{87--94}.
\newblock
\urldef\tempurl%
\url{https://doi.org/10.18178/ijke.2017.3.2.087}
\showDOI{\tempurl}


\end{thebibliography}


\end{document}